\documentclass[12pt,draftcls,onecolumn]{IEEEtran}
\usepackage{setspace}
\usepackage{amsmath,amsfonts}
\usepackage{algorithmic}
\usepackage{bbding}
\usepackage{algorithm}
\usepackage{array}
\usepackage[caption=false,font=scriptsize,labelfont=scriptsize,textfont=scriptsize]{subfig}
\usepackage{textcomp}
\usepackage{stfloats}
\usepackage{url}
\usepackage{verbatim}
\usepackage{graphicx}
\usepackage{cite}
\begin{document}

\title{Pilot Power Allocation for Channel Estimation in a Multi-RIS Aided Communication System}

\author{Jiancheng An,~\IEEEmembership{Member,~IEEE,}
and Chau Yuen,~\IEEEmembership{Fellow,~IEEE}
\thanks{This research is supported by the Ministry of Education, Singapore, under its MOE Tier 2 (Award number MOE-T2EP50220-0019) and by the Science and Engineering Research Council of A*STAR (Agency for Science, Technology and Research) Singapore, under Grant No. M22L1b0110. Any opinions, findings and conclusions or recommendations expressed in this material are those of the author(s) and do not reflect the views of the Ministry of Education or A*Star, Singapore.}
\thanks{J. An is with the Engineering Product Development (EPD) Pillar, Singapore University of Technology and Design (SUTD), Singapore 487372. (e-mail: jiancheng$\_$an@sutd.edu.sg).}
\thanks{C. Yuen is with the School of Electrical and Electronics Engineering, Nanyang Technological University, Singapore 639798 (e-mail: chau.yuen@ntu.edu.sg).}}
\markboth{IEEE GLOBECOM 2023}{IEEE GLOBECOM 2023}

\maketitle

\begin{abstract}
Reconfigurable intelligent surface (RIS) is a promising technology that enables the customization of electromagnetic propagation environments in next-generation wireless networks. In this paper, we investigate the optimal pilot power allocation during the channel estimation stage to improve the ergodic channel gain of RIS-assisted systems under practical imperfect channel state information (CSI). Specifically, we commence by deriving an explicit closed-form expression of the ergodic channel gain of a multi-RIS-aided communication system that takes into account channel estimation errors. Then, we formulate the pilot power allocation problem to maximize the ergodic channel gain under imperfect CSI, subject to the average pilot power constraint. Then, the method of Lagrange multipliers is invoked to obtain the optimal pilot power allocation solution, which indicates that allocating more power to the pilots for estimating the weak reflection channels is capable of effectively improving the ergodic channel gain under imperfect CSI. Finally, extensive simulation results corroborate our theoretical analysis.
\end{abstract}

\begin{IEEEkeywords}
Reconfigurable intelligent surface (RIS), multiple RISs, pilot power allocation, channel estimation errors.
\end{IEEEkeywords}

\section{Introduction}
\IEEEPARstart{R}{ecently}, reconfigurable intelligent surfaces (RIS) has emerged as a promising technology for constructing the future spectrum- and energy-efficient wireless networks \cite{WC_2022_An_Codebook, di2019smart, huang2019reconfigurable}. Specifically, an RIS is a programmable metasurface consisting of a huge number of passive reflecting elements, each of which is capable of altering the incident signals by imposing an independent attenuation and/or phase shift \cite{ICC_2023_An_Stacked, JSAC_2023_An_Stacked, TCOM_2021_Wu_Intelligent}. As a result, the wireless propagation environments can be customized according to the specific quality-of-service (QoS) requirements, which provides a new design degree-of-freedom for enhancing the wireless channel capacity and mitigating the multiuser interference \cite{TCOM_2022_An_Low}. In contrast to the conventional cooperative communications relying on active relays, each RIS element reflects the incident signals passively, without requiring any radio-frequency (RF) chains \cite{bjornson2019intelligent, CL_2023_An_A}. Additionally, RIS is capable of operating in the full-duplex (FD) mode without encountering severe self-interference as in the conventional FD relays \cite{TWC_2023_Xu_Antenna}. The low-cost material and patch structure of RISs make them easy to attach to environmental objects, such as walls, ceilings, and building surfaces. Thanks to these potential benefits, widespread deployment of massive RISs is expected to enable the next-generation wireless networks \cite{di2019smart, IoTJ_2023_Xu_Algorithm}.

Nevertheless, the remarkable performance gains offered by RIS heavily rely on the perfect channel state information (CSI) of all channels \cite{TCOM_2021_Wu_Intelligent, CST_2022_Zheng_A, WCL_2022_Xu_Deep}. However, the acquisition of accurate CSI, especially for the reflection channels, is much more challenging in practice due to the passive nature of RIS as well as the large number of reflecting elements it comprises \cite{WC_2022_An_Codebook}. Due to the deficiency of baseband signal processing capability, RISs are no longer able to estimate the BS-RIS link and the user-RIS link separately. To overcome this challenge, several efficient techniques have been proposed to estimate the cascaded reflection channels of the BS-RIS link and the user-RIS link instead \cite{TCOM_2021_Wu_Intelligent}. Specifically, \emph{Mishra and Johansson} \cite{mishra2019channel} have proposed an ON/OFF method to estimate the direct and cascaded reflection channels in sequence upon switching on only a single element at each time slot. By doing so, the direct channel and reflection channels can be estimated separately without incurring mutual interference. To further improve the estimation accuracy, \emph{Jensen and Carvalho} \cite{jensen2020optimal} utilized orthogonal RIS reflection patterns during the channel estimation phase, thus the cascaded reflection channels can be obtained without incurring any power loss. In addition, \emph{Wang et al.} \cite{wang2020channel} proposed a novel three-phase channel estimation approach, which estimated a set of low-dimensional scaling factors instead of directly estimating reflection channel coefficients. Furthermore, \emph{You et al.} \cite{JSAC_2020_You_Channel} reduced the pilot overhead by arranging the RIS elements into groups and estimating only a single channel coefficient for each group, at the cost of moderate performance erosion. In order to further reduce the pilot overhead, \emph{Han et al.} \cite{han2019large} utilized the statistic CSI to perform the phase shift optimization at the expense of system performance. Additionally, \emph{An et al.} \cite{TGCN_2022_An_Joint} proposed a novel codebook-based scheme to configure RIS, where only the composite channel was estimated.

Note that the aforementioned channel estimation methods \cite{mishra2019channel, jensen2020optimal, wang2020channel, JSAC_2020_You_Channel} generally utilize the constant-envelope pilot signals for each user/antenna without considering the pilot power allocation during the channel estimation phase. Furthermore, much of the prior works focus on the communication systems assisted by a single RIS, and few of them focus on the design of multi-RIS-assisted communication systems. Against the above background, in this paper, we investigate the pilot power allocation for multi-RIS-assisted wireless systems. To the best of our knowledge, this work represents the first study explored in this area. \emph{Firstly}, we present a multi-RIS-aided single-input single-output (SISO) communication system taking into account the practical channel estimation errors, where the ON/OFF method is adopted for the sake of brevity. Furthermore, we mathematically derive the ergodic channel gain under imperfect CSI in an explicit closed-form expression. \emph{Secondly}, we formulate the pilot power allocation problem to maximize the ergodic channel gain while subject to the average pilot power constraint. We then solve the formulated optimization problem by utilizing the method of Lagrange multipliers and applying some effective approximations. The resultant pilot power allocation scheme suggests allocating more power to the pilots used for estimating the weak reflection channels. \emph{Finally}, extensive simulation results verify the performance improvements of the proposed pilot power allocation scheme over the conventional average allocation counterpart.

The rest of this paper is organized as follows. Section \ref{sec2} introduces the system model. In Section \ref{sec3}, we theoretically derive the ergodic channel gain under imperfect CSI. Furthermore, Section \ref{sec4} presents the problem formulation, and Section \ref{sec5} solves the formulated problem and discusses the proposed pilot power allocation solution. In Section \ref{sec6}, extensive simulations are provided to verify our theoretical analysis. Finally, Section \ref{sec7} concludes the paper.

\emph{Notations:} Scalars are denoted by italic letters, while vectors and matrices are denoted by boldface lowercase and uppercase letters, respectively. ${\left( \cdot \right)^T}$, ${\left( \cdot \right)^H}$, and ${\left( \cdot \right)^ * }$ represent the transpose, Hermitian transpose, and conjugate operations, respectively. The space of $x \times y$ complex-valued matrices is denoted as ${\mathbb{C}^{x \times y}}$. $\text{diag}\left( {\bf{x}} \right)$ represents a diagonal matrix with each diagonal element being the corresponding element in ${\bf{x}}$. $\log \left( \cdot \right)$ is the logarithmic function, while $\mathbb{E}\left\{ \cdot \right\}$ stands for the expectation operation. The distribution of a circularly symmetric complex Gaussian (CSCG) random variable with mean $\mu$ and variance ${\sigma ^2}$ is denoted by $\mathcal{CN}\left( {\mu ,{\sigma ^2}} \right)$. $\Re \left\{ z \right\}$, $\Im \left\{ z \right\}$, and $\left| z \right|$ represent the real part, imaginary part, and magnitude of a complex number $z$, respectively.

\section{System Model}\label{sec2}
\begin{figure}[!t]
	\centering
	\includegraphics[width=9cm]{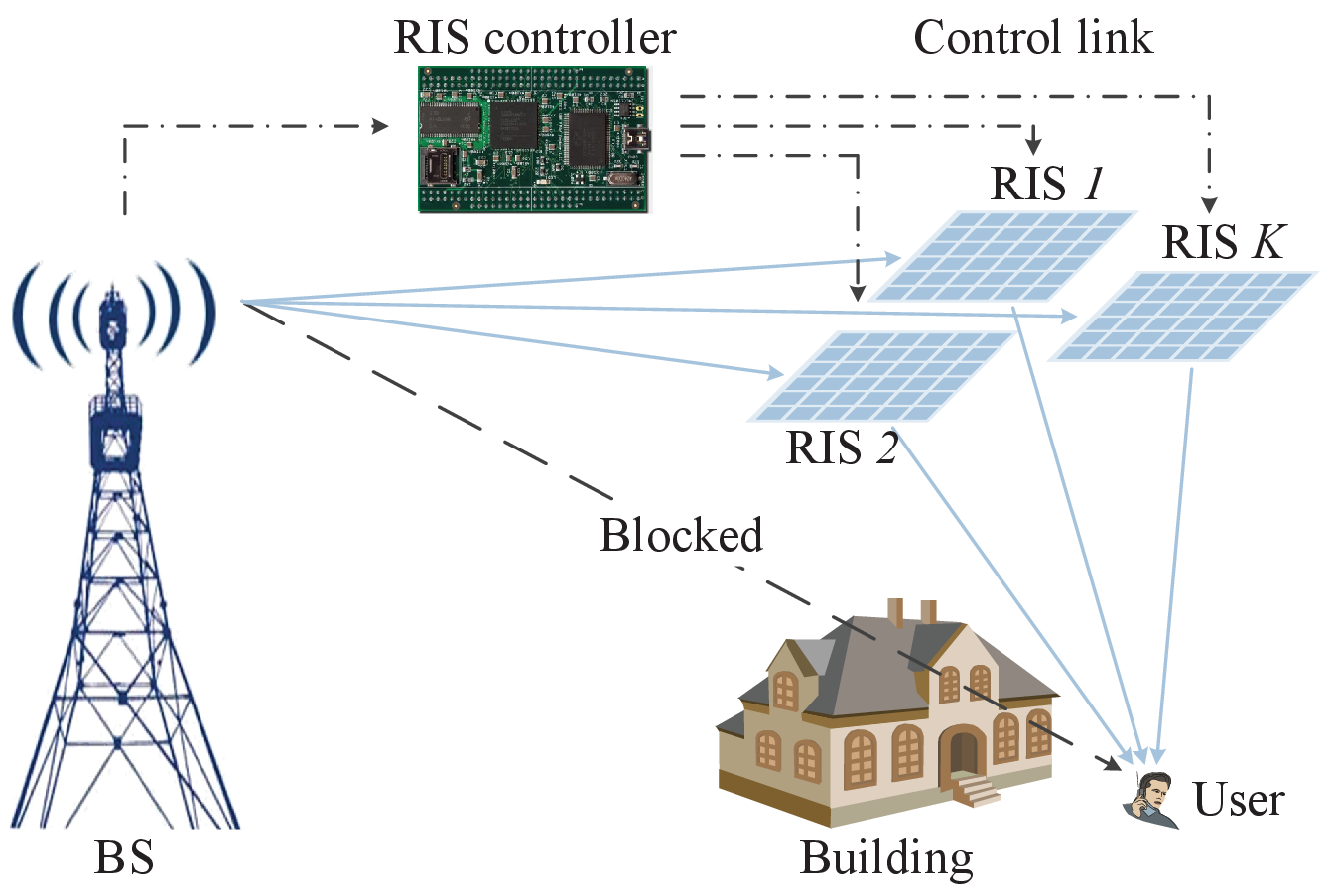}
	\caption{A multi-RIS assisted communication system.}
	\label{fig00}
\end{figure}
As shown in Fig. \ref{fig00}, we consider an RIS-assisted downlink communication system, where multiple RISs are deployed at the cell edge to enhance the communication link from the BS to the user \cite{TCOM_2021_Do_Multi}. For the sake of brevity, we consider that both the BS and the user are equipped with a single antenna\footnote{While in this paper we consider the SISO scenario, the proposed pilot power allocation scheme could readily be adapted for point-to-point and multi-user MIMO systems as well.} and that the direct link between them is blocked by a local building near the user. Let $K$ denote the number of RISs, while the set of the RISs is denoted by $\mathcal{K} = \left\{ {1, \cdots ,K} \right\}$. Furthermore, we denote the number of reflecting elements at the $k$th RIS as ${M_k},\ \forall k \in \mathcal{K}$, with the corresponding set of reflecting elements represented by ${\mathcal{M}_k} = \left\{ {1, \cdots ,{M_k}} \right\},\ \forall k \in \mathcal{K}$. All the RISs are connected to an RIS controller by the wired/wireless backhaul for conveying the reflection coefficients \cite{WC_2022_An_Codebook}. Due to the high path loss, the signals reflected by multiple RISs are assumed to be below the receiver sensitivity and are therefore ignored.

Furthermore, the quasi-static block-fading model is considered for all reflection channels. Specifically, the baseband equivalent channel from the BS to the $k$th RIS, and that from the $k$th RIS to the user are denoted by ${{\bf{u}}_k} = {\left[ {{u_{k,1}}, \cdots ,{u_{k,{M_k}}}} \right]^T}\in {\mathbb{C}^{{M_k} \times 1}}$, and ${\bf{v}}_k^H = {\left[ {{v_{k,1}}, \cdots ,{v_{k,{M_k}}}} \right]^ * }\in {\mathbb{C}^{1 \times {M_k}}}$, respectively, where ${u_{k,{m_k}}}$ and ${v_{k,{m_k}}}$ represent the channels from the BS and the user, respectively, to the $m_k$th element of the $k$th RIS. Let ${{\bf{\Phi }}_k} = \text{diag}\left( {{\varphi _{k,1}},{\varphi _{k,2}}, \cdots ,{\varphi _{k,{M_k}}}} \right) \in {\mathbb{C}^{{M_k} \times {M_k}}}$ denote the reflection coefficient matrix of the $k$th RIS, where ${{\varphi _{k,{m_k}}}}$ represents the reflection coefficient of the ${m_k}$th element of the $k$th RIS. In this paper, we focus on the phase shift design and thus assume that the magnitude at each RIS element is $\left| {{\varphi _{k,{m_k}}}} \right| = 1$ to maximize the reflection power. Furthermore, we assume that the phase shift can be continuously adjusted in $\left[ {0,2\pi } \right)$ to gain basic insight into pilot power allocation.

In this paper, we focus on narrowband communications. Denote the normalized signal transmitted from the BS over the given band as $s$, satisfying ${\left| s \right|^2} = 1$. Therefore, the baseband signal received by the user can be expressed as:
\begin{small}\begin{align}\label{eq1}
r = &\left( {\sum\limits_{k = 1}^K {\sum\limits_{{m_k} = 1}^{{M_k}} {{u_{k,{m_k}}}{\varphi _{k,{m_k}}}v_{k,{m_k}}^*} } } \right)\sqrt q s + n \notag\\
 = & \left( {\sum\limits_{k = 1}^K {\sum\limits_{{m_k} = 1}^{{M_k}} {{\varphi _{k,{m_k}}}{h_{k,{m_k}}}} } } \right)\sqrt q s + n,
\end{align}\end{small}where ${h_{k,{m_k}}} = {u_{k,{m_k}}}v_{k,{m_k}}^ * $ represents the cascaded reflection channels via the ${m_k}$th element of the $k$th RIS, $q$ is the downlink average transmit power at the BS, $n \sim \mathcal{CN}\left( {0,\sigma _n^2} \right)$ is the additive white Gaussian noise (AWGN) with $\sigma _n^2$ denoting the average noise power at the user.

Next, we consider the practical channel estimation during the uplink phase. Specifically, the ON/OFF channel estimation method is employed to estimate the cascaded reflection channels by only switching on a single reflecting element at each time slot \cite{mishra2019channel}. As such, the overall pilot overhead required is $\sum\nolimits_{k = 1}^K {{M_k}}$. Let ${x_{k,{m_k}}}$ denote the uplink pilot symbol at the $\left( {\sum\nolimits_{i = 1}^{k - 1} {{M_i}} + {m_k}} \right)$th time slot satisfying ${\left| {{x_{k,{m_k}}}} \right|^2} = 1$, which is used for estimating the reflection channel via the $m_k$th element of the $k$th RIS. Therefore, the uplink channel estimation model at the $\left( {\sum\nolimits_{i = 1}^{k - 1} {{M_i}} + {m_k}} \right)$th time slot can be formulated by
\begin{small}\begin{align}\label{eq3}
 {y_{k,{m_k}}} = {\left( {{\varphi _{k,{m_k}}}{h_{k,{m_k}}}} \right)^ * }\sqrt {{p_{k,{m_k}}}} {x_{k,{m_k}}} + {z_{k,{m_k}}},
\end{align}\end{small}where ${p_{k,{m_k}}}$ is the corresponding uplink pilot power, ${z_{k,{m_k}}} \sim \mathcal{CN}\left( {0,\sigma _z^2} \right)$ is the AWGN at the BS, ${\varphi _{k,{m_k}}}$ is the RIS reflection coefficient at the current time slot, which can be set to be an arbitrary complex value satisfying $\left| {{\varphi _{k,{m_k}}}} \right| = 1$.

According to \eqref{eq3}, the least squares (LS) estimate of the cascaded reflection channel via the $m_k$th element of the $k$th RIS can be obtained by
\begin{small}\begin{align}\label{eq4}
{{\hat h}_{k,{m_k}}} = \frac{1}{{\sqrt {{p_{k,{m_k}}}} }}y_{k,{m_k}}^ * {x_{k,{m_k}}}\varphi _{k,{m_k}}^ * = {h_{k,{m_k}}} + {\varepsilon _{k,{m_k}}},
\end{align}\end{small}where ${\varepsilon _{k,{m_k}}} = \frac{1}{{\sqrt {{p_{k,{m_k}}}} }}z_{k,{m_k}}^*{x_{k,{m_k}}}\varphi _{k,{m_k}}^*$ denotes the estimation error with respect to ${h_{k,{m_k}}}$. It is evident that we have ${\varepsilon _{k,{m_k}}} \sim \mathcal{CN}\left( {0,\delta _{k,{m_k}}^2} \right)$ with $\delta _{k,{m_k}}^2 = {{{\sigma _z^2} \mathord{\left/
 {\vphantom {{\sigma _z^2} {{p_{k,{m_k}}}}}} \right.
 \kern-\nulldelimiterspace} {{p_{k,{m_k}}}}}}$ denoting the mean square error (MSE) of the LS channel estimate.

Based on the estimated ${\hat h_{k,{m_k}}}$ in \eqref{eq4}, the RIS controller will adapt the reflection coefficient of the $m_k$th element of the $k$th RIS to
\begin{small}\begin{align}\label{eq5}
{\hat \varphi _{k,{m_k}}} = \frac{{\hat h_{k,{m_k}}^*}}{{\left| {{{\hat h}_{k,{m_k}}}} \right|}} = \frac{{h_{k,{m_k}}^* + \varepsilon _{k,{m_k}}^*}}{{\left| {{h_{k,{m_k}}} + {\varepsilon _{k,{m_k}}}} \right|}}.
\end{align}\end{small}

Thus, the actual composite channel $h$ between the BS and the user is formulated by
\begin{small}\begin{align}\label{eq8}
h = \sum\limits_{k = 1}^K {\sum\limits_{{m_k} = 1}^{{M_k}} {{{\hat \varphi }_{k,{m_k}}}{h_{k,{m_k}}}} }.
\end{align}\end{small}

\section{Ergodic  Channel Gain under Imperfect CSI}\label{sec3}
In this section, we will derive the explicit closed-form expression of the ergodic channel gain of the multi-RIS-assisted SISO system under imperfect CSI, i.e., $\mathbb{E}\left\{ {{{\left| h \right|}^2}} \right\}$. Upon substituting \eqref{eq5} into \eqref{eq8}, the ergodic channel gain can be expressed by \eqref{eq9},
\begin{figure*}[!t]
\begin{small}\begin{align}\label{eq9}
\mathbb{E}\left\{ {{{\left| h \right|}^2}} \right\} =& \mathbb{E}\left\{ {{{\left| {\sum\limits_{k = 1}^K {\sum\limits_{{m_k} = 1}^{{M_k}} {\frac{{\left( {h_{k,{m_k}}^ * + \varepsilon _{k,{m_k}}^ * } \right){h_{k,{m_k}}}}}{{\left| {{h_{k,{m_k}}} + {\varepsilon _{k,{m_k}}}} \right|}}} } } \right|}^2}} \right\} \overset{\left( a \right)}{=} \sum\limits_{k = 1}^K {\sum\limits_{{m_k} = 1}^{{M_k}} {\mathbb{E}\left\{ {{{\left| {{h_{k,{m_k}}}} \right|}^2}} \right\}} } \notag\\
+& \sum\limits_{k = 1}^K {\sum\limits_{{m_k} = 1}^{{M_k}} {\mathbb{E}\left\{ {\Re \left\{ {\frac{{\left( {h_{k,{m_k}}^ * + \varepsilon _{k,{m_k}}^ * } \right){h_{k,{m_k}}}}}{{\left| {{h_{k,{m_k}}} + {\varepsilon _{k,{m_k}}}} \right|}}} \right\}} \right\}} } \sum\limits_{{{\tilde m}_k} = 1,{{\tilde m}_k} \ne {m_k}}^{{M_k}} {\mathbb{E}\left\{ {\Re \left\{ {\frac{{h_{k,{{\tilde m}_k}}^ * + \varepsilon _{k,{{\tilde m}_k}}^ * {h_{k,{{\tilde m}_k}}}}}{{\left| {{h_{k,{{\tilde m}_k}}} + {\varepsilon _{k,{{\tilde m}_k}}}} \right|}}} \right\}} \right\}} \notag\\
 +& \sum\limits_{k = 1}^K {\sum\limits_{{m_k} = 1}^{{M_k}} {\mathbb{E}\left\{ {\Im \left\{ {\frac{{h_{k,{m_k}}^ * + \varepsilon _{k,{m_k}}^ * {h_{k,{m_k}}}}}{{\left| {{h_{k,{m_k}}} + {\varepsilon _{k,{m_k}}}} \right|}}} \right\}} \right\}} } \sum\limits_{{{\tilde m}_k} = 1,{{\tilde m}_k} \ne {m_k}}^{{M_k}} {\mathbb{E}\left\{ {\Im \left\{ {\frac{{h_{k,{{\tilde m}_k}}^ * + \varepsilon _{k,{{\tilde m}_k}}^ * {h_{k,{{\tilde m}_k}}}}}{{\left| {{h_{k,{{\tilde m}_k}}} + {\varepsilon _{k,{{\tilde m}_k}}}} \right|}}} \right\}} \right\}} \notag\\
 +& \sum\limits_{k = 1}^K {\sum\limits_{{m_k} = 1}^{{M_k}} {\mathbb{E}\left\{ {\Re \left\{ {\frac{{\left( {h_{k,{m_k}}^ * + \varepsilon _{k,{m_k}}^ * } \right){h_{k,{m_k}}}}}{{\left| {{h_{k,{m_k}}} + {\varepsilon _{k,{m_k}}}} \right|}}} \right\}} \right\}} } \sum\limits_{\tilde k = 1,\tilde k \ne k}^K {\sum\limits_{{m_{\tilde k}} = 1}^{{M_{\tilde k}}} {\mathbb{E}\left\{ {\Re \left\{ {\frac{{\left( {h_{\tilde k,{m_{\tilde k}}}^* + \varepsilon _{\tilde k,{m_{\tilde k}}}^*} \right){h_{\tilde k,{m_{\tilde k}}}}}}{{\left| {{h_{\tilde k,{m_{\tilde k}}}} + {\varepsilon _{\tilde k,{m_{\tilde k}}}}} \right|}}} \right\}} \right\}} } \notag\\
 +& \sum\limits_{k = 1}^K {\sum\limits_{{m_k} = 1}^{{M_k}} {\mathbb{E}\left\{ {\Im \left\{ {\frac{{\left( {h_{k,{m_k}}^ * + \varepsilon _{k,{m_k}}^ * } \right){h_{k,{m_k}}}}}{{\left| {{h_{k,{m_k}}} + {\varepsilon _{k,{m_k}}}} \right|}}} \right\}} \right\}} } \sum\limits_{\tilde k = 1,\tilde k \ne k}^K {\sum\limits_{{m_{\tilde k}} = 1}^{{M_{\tilde k}}} {\mathbb{E}\left\{ {\Im \left\{ {\frac{{\left( {h_{\tilde k,{m_{\tilde k}}}^* + \varepsilon _{\tilde k,{m_{\tilde k}}}^*} \right){h_{\tilde k,{m_{\tilde k}}}}}}{{\left| {{h_{\tilde k,{m_{\tilde k}}}} + {\varepsilon _{\tilde k,{m_{\tilde k}}}}} \right|}}} \right\}} \right\}} }.
\end{align}\end{small}
\hrulefill
\end{figure*}
where $\left( a \right)$ holds as the reflection channel ${h_{k,{m_k}}}$ and the estimation error ${\varepsilon _{k,{m_k}}}$ associated with different reflecting elements are independent of each other. Furthermore, the second and third terms on the right-hand side of \eqref{eq9} characterize the impacts of the channel estimation errors from the same RIS, while the fourth and fifth terms depict the impacts of the channel estimation errors from the different RISs.

Furthermore, in the practical implementations, the BS and RIS are generally installed at high-altitude locations with a small number of local scatters, resulting in a line-of-sight (LoS) propagation link with a high Rician factor and long channel coherence time \cite{WC_2022_An_Codebook}. Therefore, it is reasonable to assume that the small-scale fading of the cascaded reflection channels is predominantly determined by the RIS-user link, i,e., ${h_{k,{m_k}}} \sim \mathcal{CN}\left( {0,\beta _k^2} \right)$, where $\beta _k^2$ denotes the path loss of the cascaded reflection channels through the $k$th RIS. Note that all the ${M_k}$ elements within the $k$th RIS experience the same degree of path loss. As a result, for the $m_k$th reflecting element of the $k$th RIS, we have $\mathbb{E}\left\{ {{{\left| {{h_{k,{m_k}}}} \right|}^2}} \right\} = \beta _k^2$. Additionally, according to \emph{Proposition 4} of \cite{TGCN_2022_An_Joint}, we have
\begin{small}\begin{align}
\mathbb{E}\left\{ {\Re \left\{ {\frac{{\left( {h_{k,{m_k}}^* + \varepsilon _{k,{m_k}}^*} \right){h_{k,{m_k}}}}}{{\left| {{h_{k,{m_k}}} + {\varepsilon _{k,{m_k}}}} \right|}}} \right\}} \right\} =& \frac{{\sqrt \pi \beta _k^2}}{{2\sqrt {\beta _k^2 + \delta _{k,{m_k}}^2} }}, \label{eq10}\\
\mathbb{E}\left\{ {\Im \left\{ {\frac{{\left( {h_{k,{m_k}}^* + \varepsilon _{k,{m_k}}^*} \right){h_{k,{m_k}}}}}{{\left| {{h_{k,{m_k}}} + {\varepsilon _{k,{m_k}}}} \right|}}} \right\}} \right\} =& 0. \label{eq11}
\end{align}\end{small}

Upon substituting \eqref{eq10} and \eqref{eq11} into \eqref{eq9}, the ergodic channel gain can be further simplified as shown in \eqref{eq12}.
\begin{figure*}
\begin{small}\begin{align}
\mathbb{E}\left\{ {{{\left| h \right|}^2}} \right\} &= \sum\limits_{k = 1}^K {{M_k}\beta _k^2} +\frac{\pi }{4}\sum\limits_{k = 1}^K {\beta _k^4\sum\limits_{{m_k} = 1}^{{M_k}} {\frac{1}{{\sqrt {\beta _k^2 + \frac{{\sigma _z^2}}{{{p_{k,{m_k}}}}}} }}\sum\limits_{{{\tilde m}_k} = 1,{{\tilde m}_k} \ne {m_k}}^{{M_k}} {\frac{1}{{\sqrt {\beta _k^2 + \frac{{\sigma _z^2}}{{{p_{k,{{\tilde m}_k}}}}}} }}} } } \notag\\
&+ \frac{\pi }{4}\sum\limits_{k = 1}^K {\sum\limits_{{m_k} = 1}^{{M_k}} {\frac{{\beta _k^2}}{{\sqrt {\beta _k^2 + \frac{{\sigma _z^2}}{{{p_{k,{m_k}}}}}} }}\sum\limits_{\tilde k = 1,\tilde k \ne k}^K {\sum\limits_{{m_{\tilde k}} = 1}^{{M_{\tilde k}}} {\frac{{\beta _{\tilde k}^2}}{{\sqrt {\beta _{\tilde k}^2 + \frac{{\sigma _z^2}}{{{p_{\tilde k,{m_{\tilde k}}}}}}} }}} } } }. \label{eq12}
\end{align}\end{small}
\hrulefill
\end{figure*}
We note that the ergodic channel gain in \eqref{eq12} depends on the power of pilots used for estimating different reflection channels. As such, the ergodic channel gain under imperfect CSI can be further improved by appropriately allocating the pilot power.
\section{Problem Formulation of the Pilot Power Allocation}\label{sec4}
In this section, we will provide the problem formulation for optimizing the pilot power to maximize the ergodic channel gain under imperfect CSI. After removing the constant terms and positive coefficient, the optimization problem is expressed as 
\begin{small}\begin{align}\label{eq14}
\begin{array}{*{20}{l}}
{\left( {P_\mathcal{A}} \right)}&{\mathop {\max }\limits_\mathcal{P} }&{\varphi \left( \mathcal{P} \right)}\\
{}&{\text{s.t.}}&{\sum\limits_{k = 1}^K {\sum\limits_{{m_k} = 1}^{{M_k}} {{p_{k,{m_k}}}} } = \sum\limits_{k = 1}^K {{M_k}} p},
\end{array}
\end{align}\end{small}where $p$ is the average uplink pilot power at the user; while $\mathcal{P} = \left\{ {{p_{1,1}}, \cdots ,{p_{1,{M_1}}}, \cdots ,{p_{K,1}}, \cdots ,{p_{K,{M_K}}}} \right\}$ is the set of the specific pilot power for entire $\sum\nolimits_{k = 1}^K {{M_k}} $ time slots during the channel estimation phase, where the equivalent objective function $\varphi \left( \mathcal{P} \right)$ is defined by \eqref{eq_new11}.
\begin{figure*}
\begin{small}\begin{align}\label{eq_new11}
\varphi \left( \mathcal{P} \right) = \sum\limits_{k = 1}^K {\beta _k^4\sum\limits_{{m_k} = 1}^{{M_k}} {\frac{1}{{\sqrt {\beta _k^2 + \frac{{\sigma _z^2}}{{{p_{k,{m_k}}}}}} }}} \sum\limits_{{{\tilde m}_k} = 1,{{\tilde m}_k} \ne {m_k}}^{{M_k}} {\frac{1}{{\sqrt {\beta _k^2 + \frac{{\sigma _z^2}}{{{p_{k,{{\tilde m}_k}}}}}} }}} } + \sum\limits_{k = 1}^K {\sum\limits_{{m_k} = 1}^{{M_k}} {\frac{{\beta _k^2}}{{\sqrt {\beta _k^2 + \frac{{\sigma _z^2}}{{{p_{k,{m_k}}}}}} }}} } \sum\limits_{\tilde k = 1,\tilde k \ne k}^K {\sum\limits_{{m_{\tilde k}} = 1}^{{M_{\tilde k}}} {\frac{{\beta _{\tilde k}^2}}{{\sqrt {\beta _{\tilde k}^2 + \frac{{\sigma _z^2}}{{{p_{\tilde k,{m_{\tilde k}}}}}}} }}} }.
\end{align}\end{small}
\hrulefill
\end{figure*}
\section{The Optimal Pilot Power Allocation Scheme}\label{sec5}
In this section, we will utilize the method of Lagrange multipliers to derive the solution of the problem $\left( {P_\mathcal{A}} \right)$. To begin, the Lagrangian function $f\left( {\mathcal{P},\lambda } \right)$ is defined as follows
\begin{small}\begin{align}
f\left( {\mathcal{P},\lambda } \right) = \varphi \left( \mathcal{P} \right) - \lambda \left( {\sum\limits_{k = 1}^K {\sum\limits_{{m_k} = 1}^{{M_k}} {{p_{k,{m_k}}}} } - \sum\limits_{k = 1}^K {{M_k}} p} \right),
\end{align}\end{small}where $\lambda $ is the Lagrange multiplier.

Next, we proceed to determine the stationary point of the Lagrange function upon setting the first-order partial derivatives of $f\left( {\mathcal{P},\lambda } \right)$ to zero. More specifically, for the $m_k$th element of the $k$th RIS, we arrive at \eqref{eq17} as shown on the top of the next page.
\begin{figure*}
\begin{small}\begin{align}\label{eq17}
\frac{{\partial f}}{{\partial {p_{k,{m_k}}}}} = \frac{{\beta _k^4\sigma _z^2}}{{p_{k,{m_k}}^2\sqrt {{{\left( {\beta _k^2 + \frac{{\sigma _z^2}}{{{p_{k,{m_k}}}}}} \right)}^3}} }}\sum\limits_{{{\tilde m}_k} = 1,{{\tilde m}_k} \ne {m_k}}^{{M_k}} {\frac{1}{{\sqrt {\beta _k^2 + \frac{{\sigma _z^2}}{{{p_{k,{{\tilde m}_k}}}}}} }}} + \frac{{\beta _k^2\sigma _z^2}}{{p_{k,{m_k}}^2\sqrt {{{\left( {\beta _k^2 + \frac{{\sigma _z^2}}{{{p_{k,{m_k}}}}}} \right)}^3}} }}\sum\limits_{\tilde k = 1,\tilde k \ne k}^K {\sum\limits_{{m_{\tilde k}} = 1}^{{M_{\tilde k}}} {\frac{{\beta _{\tilde k}^2}}{{\sqrt {\beta _{\tilde k}^2 + \frac{{\sigma _z^2}}{{{p_{\tilde k,{m_{\tilde k}}}}}}} }}} } - \lambda =0.
\end{align}\end{small}
\hrulefill
\end{figure*}
Furthermore, let us focus on the set of the first-order partial derivatives associated with the $k$th RIS, i.e., $\left\{ {{{\partial f} \mathord{\left/
 {\vphantom {{\partial f} {\partial {p_{k,1}}}}} \right.
 \kern-\nulldelimiterspace} {\partial {p_{k,1}}}}, \cdots ,{{\partial f} \mathord{\left/
 {\vphantom {{\partial f} {\partial {p_{k,{M_k}}}}}} \right.
 \kern-\nulldelimiterspace} {\partial {p_{k,{M_k}}}}}} \right\}$. By comparing arbitrary pairs of $\left( {{m_k} \in {\mathcal{M}_k},\ {m'_k} \in {\mathcal{M}_k}} \right)$ and constructing identical terms on both sides, we have \eqref{eq19},
 \begin{figure*}[!t]
\begin{small}\begin{align}\label{eq19}
\frac{{\beta _k^2\left( {{A_k} - {1 \mathord{\left/
 {\vphantom {1 {\sqrt {\beta _k^2 + \frac{{\sigma _z^2}}{{{p_{k,{m_k}}}}}} }}} \right.
 \kern-\nulldelimiterspace} {\sqrt {\beta _k^2 + \frac{{\sigma _z^2}}{{{p_{k,{m_k}}}}}} }}} \right)}}{{p_{k,{m_k}}^2\sqrt {{{\left( {\beta _k^2 + \frac{{\sigma _z^2}}{{{p_{k,{m_k}}}}}} \right)}^3}} }} + \frac{{{B_k}}}{{p_{k,{m_k}}^2\sqrt {{{\left( {\beta _k^2 + \frac{{\sigma _z^2}}{{{p_{k,{m_k}}}}}} \right)}^3}} }} = \frac{{\beta _k^2\left( {{A_k} - {1 \mathord{\left/
 {\vphantom {1 {\sqrt {\beta _k^2 + \frac{{\sigma _z^2}}{{{p_{k,{m'_k}}}}}} }}} \right.
 \kern-\nulldelimiterspace} {\sqrt {\beta _k^2 + \frac{{\sigma _z^2}}{{{p_{k,{m'_k}}}}}} }}} \right)}}{{p_{k,{m'_k}}^2\sqrt {{{\left( {\beta _k^2 + \frac{{\sigma _z^2}}{{{p_{k,{m'_k}}}}}} \right)}^3}} }} + \frac{{{B_k}}}{{p_{k,{m'_k}}^2\sqrt {{{\left( {\beta _k^2 + \frac{{\sigma _z^2}}{{{p_{k,{m'_k}}}}}} \right)}^3}} }}.
\end{align}\end{small}
\hrulefill
 \end{figure*}
where ${A_k}$ and ${B_k}$ are defined by
\begin{small}\begin{align}
{A_k} =& \sum\limits_{{m_k} = 1}^{{M_k}} {\frac{1}{{\sqrt {\beta _k^2 + \frac{{\sigma _z^2}}{{{p_{k,{m_k}}}}}} }}},\\
{B_k} =& \sum\limits_{\tilde k = 1,\tilde k \ne k}^K {\sum\limits_{{m_{\tilde k}} = 1}^{{M_{\tilde k}}} {\frac{{\beta _{\tilde k}^2}}{{\sqrt {\beta _{\tilde k}^2 + \frac{{\sigma _z^2}}{{{p_{\tilde k,{m_{\tilde k}}}}}}} }}} },
\end{align}\end{small}respectively, which are independent of ${m_k}$ and ${m'_k}$.

Based on \eqref{eq19}, it is evident that ${p_{k,{m_k}}} = {p_{k,{m'_k}}}$ for $\forall {m_k} \in {\mathcal{M}_k},\ \forall {m'_k} \in {\mathcal{M}_k}$. As a result, the pilot power for estimating the cascaded reflection channels associated with the same RIS can be determined by
\begin{small}\begin{align}\label{eq22}
{p_{k,1}} = \cdots = {p_{k,{M_k}}} = {p_k},\quad \forall k \in \mathcal{K},
\end{align}\end{small}where ${p_k}$ represents the average pilot power allocated for the $k$th RIS. It is implied that an equal amount of power should be allocated to the pilots for estimating the channel coefficients of the reflection link through the same RIS, which is consistent with our intuition as these channels experience the same large-scale fading. In the next, we will concentrate on the pilot power allocation among different RISs that experience different levels of path loss.

Specifically, by substituting \eqref{eq22} into \eqref{eq17} and simplifying it, we have
\begin{small}\begin{align}\label{eq23}
\frac{{\beta _k^2\sigma _z^2}}{{p_k^2\sqrt {{{\left( {\beta _k^2 + \frac{{\sigma _z^2}}{{{p_k}}}} \right)}^3}} }}\sum\limits_{\tilde k = 1}^K {\frac{{\beta _{\tilde k}^2{M_{\tilde k}}}}{{\sqrt {\left( {\beta _{\tilde k}^2 + \frac{{\sigma _z^2}}{{{p_{\tilde k}}}}} \right)} }}} - \frac{{\beta _k^4\sigma _z^2}}{{p_k^2{{\left( {\beta _k^2 + \frac{{\sigma _z^2}}{{{p_k}}}} \right)}^2}}} - \lambda = 0,
\end{align}\end{small}for $k \in \mathcal{K}$.

Nevertheless, it is non-trivial to solve the set of $K$ equations in \eqref{eq23}, since they are actually higher-degree equations with respect to $\left\{ {{p_1}, \cdots ,{p_K}} \right\}$ and are coupled with each other. To address this issue, we will find a high-quality approximate solution instead. Considering that in the practical implementations, communication systems generally operate at a moderate SNR region to satisfy the specific QoS requirement, i.e., ${{{p_k}\beta _k^2} \mathord{\left/
 {\vphantom {{{p_k}\beta _k^2} {\sigma _z^2}}} \right.
 \kern-\nulldelimiterspace} {\sigma _z^2}} \ge \gamma $, where $\gamma $ is the minimum SNR requirement. Consequently, it is reasonable to assume that $\beta _k^2 + {{\sigma _z^2} \mathord{\left/
 {\vphantom {{\sigma _z^2} {{p_k}}}} \right.
 \kern-\nulldelimiterspace} {{p_k}}} \approx \beta _k^2,\ \forall k \in \mathcal{K}$. By applying such an approximation, \eqref{eq23} can be simplified to
\begin{small}\begin{align}
\frac{{\sigma _z^2}}{{p_k^2{\beta _k}}}\sum\limits_{\tilde k = 1}^K {{\beta _{\tilde k}}{M_{\tilde k}}} - \frac{{\sigma _z^2}}{{p_k^2}} - \lambda = 0.
 \end{align}\end{small}
 
Furthermore, recalling that the average pilot power constraint of $\sum\limits_{k = 1}^K {{p_k}{M_k}} = \sum\limits_{k = 1}^K {{M_k}} p$, the optimal pilot power allocation solution can be obtained by
\begin{small}\begin{align}\label{eq27}
{p_k} = \frac{{\sqrt {\left( {\frac{1}{{{\beta _k}}}\sum\limits_{\hat k = 1}^K {{\beta _{\hat k}}{M_{\hat k}}} - 1} \right)} }}{{\sum\limits_{\tilde k = 1}^K {{M_{\tilde k}}\sqrt {\left( {\frac{1}{{{\beta _{\tilde k}}}}\sum\limits_{\hat k = 1}^K {{\beta _{\hat k}}{M_{\hat k}}} - 1} \right)} } }}\sum\limits_{\tilde k = 1}^K {{M_{\tilde k}}} p.
\end{align}\end{small}

In the practical deployment, each RIS is generally equipped with a large number of reflecting elements to compensate for the high path loss. As a result, we have $\frac{1}{{{\beta _k}}}\sum\nolimits_{\hat k = 1}^K {{\beta _{\hat k}}{M_{\hat k}}} \gg 1$ for $\forall k \in \mathcal{K}$. Accordingly, the power allocation scheme in \eqref{eq27} can be further simplified to
\begin{small}\begin{align}\label{eq28}
{p_k} = \frac{{\sum\limits_{\tilde k = 1}^K {{M_{\tilde k}}} p}}{{\sqrt {{\beta _k}} \sum\limits_{\tilde k = 1}^K {\frac{{{M_{\tilde k}}}}{{\sqrt {{\beta _{\tilde k}}} }}} }}.
\end{align}\end{small}

When considering the scenario where all RISs are equipped with an equal number of reflecting elements, i.e., ${M_1} = \cdots = {M_K}$, the optimal pilot power allocation solution in \eqref{eq28} is reduced to
\begin{small}\begin{align}\label{eq28-1}
{p_k} = \frac{{Kp}}{{\sqrt {{\beta _k}} \sum\limits_{\tilde k = 1}^K {\frac{1}{{\sqrt {{\beta _{\tilde k}}} }}} }},
\end{align}\end{small}which characterizes the inverse square-root pilot power scaling law with respect to the path loss of the corresponding reflection channel, i.e., ${p_k} \propto {1 \mathord{\left/
 {\vphantom {1 {\sqrt {{\beta _k}} }}} \right.
 \kern-\nulldelimiterspace} {\sqrt {{\beta _k}} }}$.
 \begin{figure}[!t]
	\centering
	\includegraphics[width=9cm]{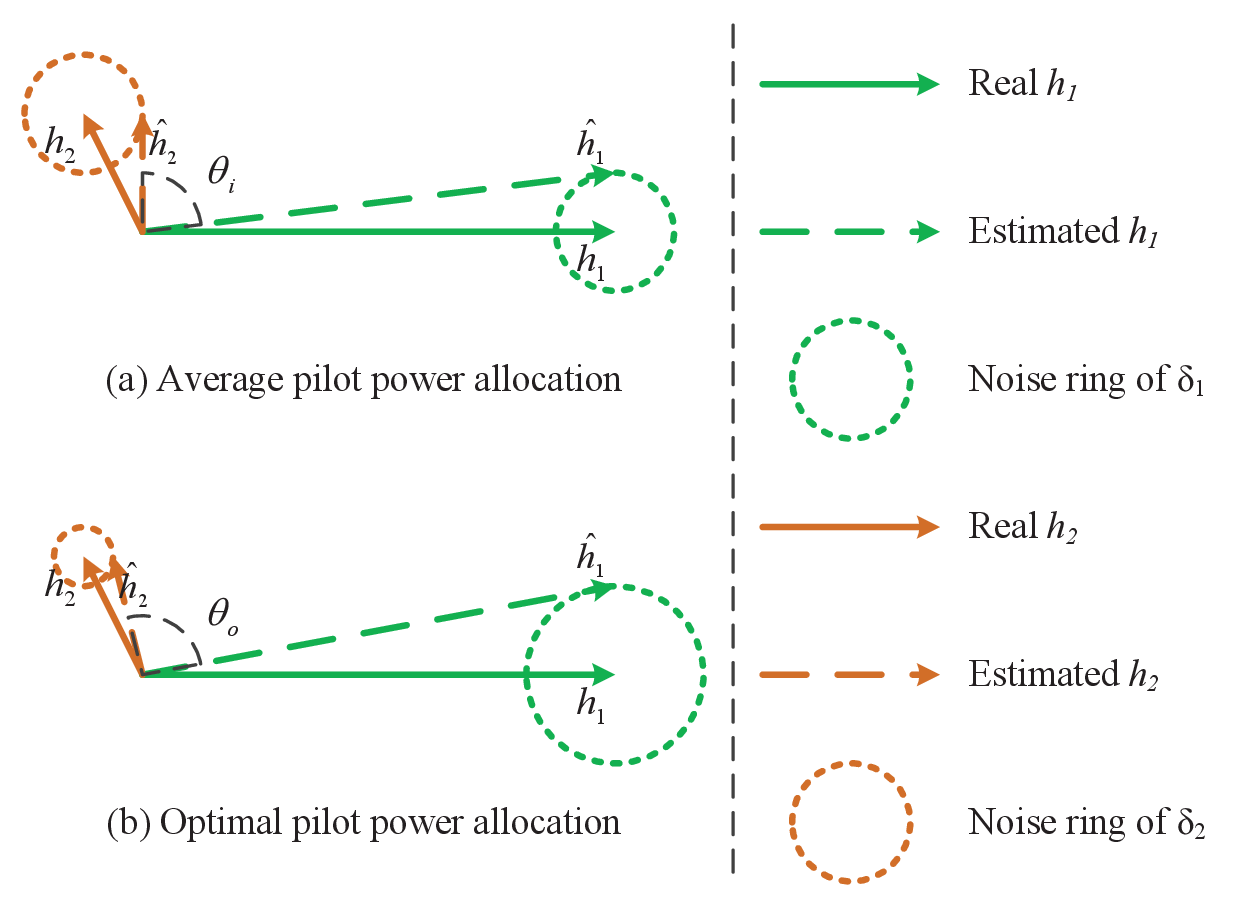}
	\caption{Comparison of two pilot power allocation schemes, where we have $K = 2$, ${M_1} = {M_2}=1$, and $\beta _1^2 = 16\beta _2^2$.}
	\label{fig2}
\end{figure}

\emph{Explicitly, \eqref{eq28} and \eqref{eq28-1} imply that one should allocate more power to obtain a more accurate estimate of the weak reflection channels.} To explain the conclusions in \eqref{eq28-1} in more detail, we provide a simple example in Fig. \ref{fig2}, where a communication system assisted by two RISs is considered. Each RIS is equipped with a single reflecting element, and the statistical channel gain of such two links is set such that ${\beta _1^2} = 16{\beta _2^2}$. \emph{It should be noted that the phase diffusion of the channel estimate is more severe for the weak channel than for the strong channel. Therefore, attaining a more accurate estimate of the weak channel at the expense of a less accurate estimate of the strong channel is a compromise.} As illustrated in Fig. \ref{fig2}, the proposed optimal pilot power allocation scheme is capable of improving the accuracy of the RIS phase shift in the statistical sense.

\section{Simulation Results}\label{sec6}
In this section, we provide simulation results to validate our analysis. The average noise power at the BS and user are set to $\sigma _z^2 = - 110$ dBm and $\sigma _n^2 = - 90$ dBm, respectively. The average downlink transmit power at the BS is set to $q = 40$ dBm. In our simulations, the BS is located at $\left( {0,0,{d_h}} \right)$ with the height of ${d_h} = 10$ m. We assume that the direct link between the BS and the user is blocked during the user's movement. The path loss of the BS-RIS and RIS-user links is modeled by ${\beta ^2} = {C_0}{d^{ - \alpha }}$, where ${C_0} = - 20$ dB represents the path loss at the reference distance of ${d_r} = 1$ m. The path loss exponents of the BS-RIS and RIS-user links are set to ${\alpha _{br}} = 2.2$ and ${\alpha _{ru}} = 2.8$, respectively \cite{TCOM_2022_An_Low}. Furthermore, the Rician fading channel model is adopted for both the BS-RIS link and the RIS-user link, with the Rician factors of ${K_{br}} = \infty $ and ${K_{ru}} = 0$, respectively. All the simulation results are obtained by averaging $1,000$ independent channel realizations. For the sake of brevity, other system parameters will be specified in the following figures.

\begin{figure}[!t]
	\centering
	\includegraphics[width=9cm]{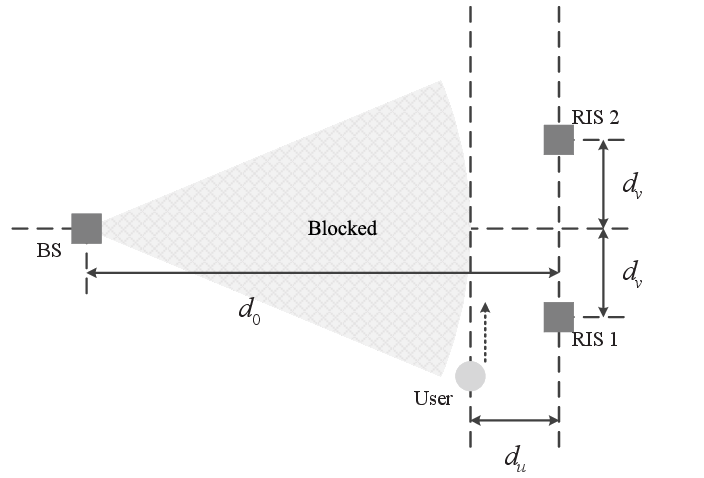}
	\caption{The simulation setup of the two-RIS-assisted communication system (view from the top).}
	\label{fig5}
\end{figure}

\begin{figure}[!t]
\centering
\subfloat[The ergodic achievable rate versus the $y$-coordinate of the user.]{\includegraphics[width=8.5cm]{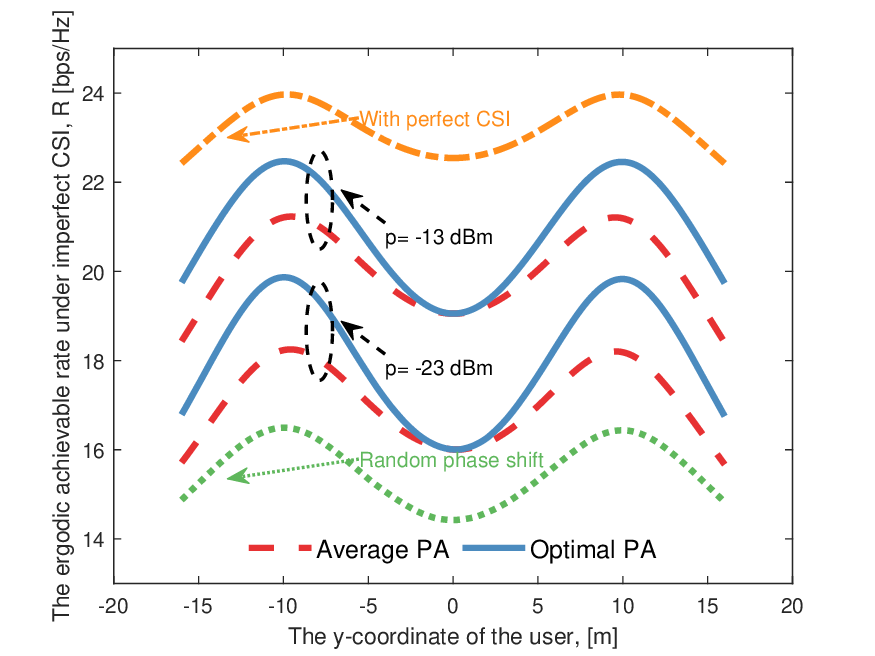}%
\label{fig6}}
\subfloat[The power of pilot signals versus the $y$-coordinate of the user.]{\includegraphics[width=8.5cm]{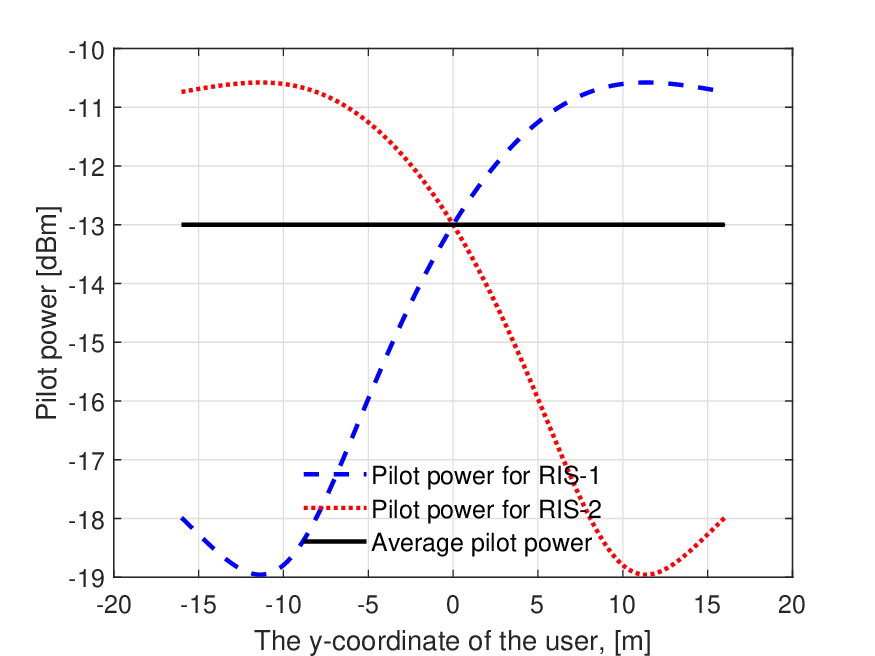}%
\label{fig7}}
\caption{Symmetrical arrangement of two RISs.}
\end{figure}

Furthermore, we consider a communication system with two distributed RISs to elaborate on the proposed pilot power allocation scheme. As shown in Fig. \ref{fig5}, RIS-$1$ and RIS-$2$ are located at $\left( {{d_0}, - {d_v},{d_h}} \right)$ and $\left( {{d_0},{d_v},{d_h}} \right)$, respectively, where ${d_v}$ is set to ${d_v} = 10$ m. The user is located at the line parallel to the line connecting RIS-$1$ and RIS-$2$, with the horizontal distance of ${d_u} = 2$ m. As such, the user's location is denoted by $\left( {{d_0} - {d_u},d,0} \right)$. The number of reflecting elements on RIS-$1$ and RIS-$2$ is set to ${M_1} = {M_2} = 100$. For the sake of illustration, Fig. \ref{fig6} evaluates the ergodic achievable rate for different power allocation strategies, which relies on the ergodic channel gain. The $y$-coordinate of the user is varied from $d = - 16$ m to $d = 16$ m. To elaborate, the ergodic achievable rate under the optimal phase shift configuration with perfect CSI and that with a random phase shift configuration are also plotted in Fig. \ref{fig6}. Specifically, we consider two cases of the average pilot power of $p = - 13$ dBm and $p = - 23$ dBm, respectively. While the increased average pilot power leads to performance improvements, the proposed pilot power allocation scheme significantly widens the performance gap with the average pilot power allocation scheme, especially for the user in the vicinity of RISs. This promising result is consistent with our theoretical analysis, indicating that appropriately allocating the power levels of the pilots used for estimating different reflection channels can improve the ergodic achievable rate under imperfect CSI.

Moreover, the specific pilot power used for estimating the reflection channels of the BS--RIS-$1$--user link and the BS--RIS-$2$--user link are shown in Fig. \ref{fig7}, with the average pilot power of $p = - 13$ dBm. It can be seen from Fig. \ref{fig7} that when the user moves in the vicinity of RIS-$1$, the proposed pilot power allocation scheme suggests allocating more power to the pilots for estimating the reflection channels via RIS-$2$, which is somewhat counter-intuitive but effectively improves the achievable rate of the user near RIS-$1$. Similar results can be observed when the user moves closer to RIS-$2$. Specifically, when the user is in the middle of RIS-$1$ and RIS-$2$, the proposed pilot power allocation scheme is equivalent to the average pilot power allocation scheme.

\begin{figure}[!t]
\centering
\subfloat[The ergodic achievable rate versus the $y$-coordinate of the user.]{\includegraphics[width=8.5cm]{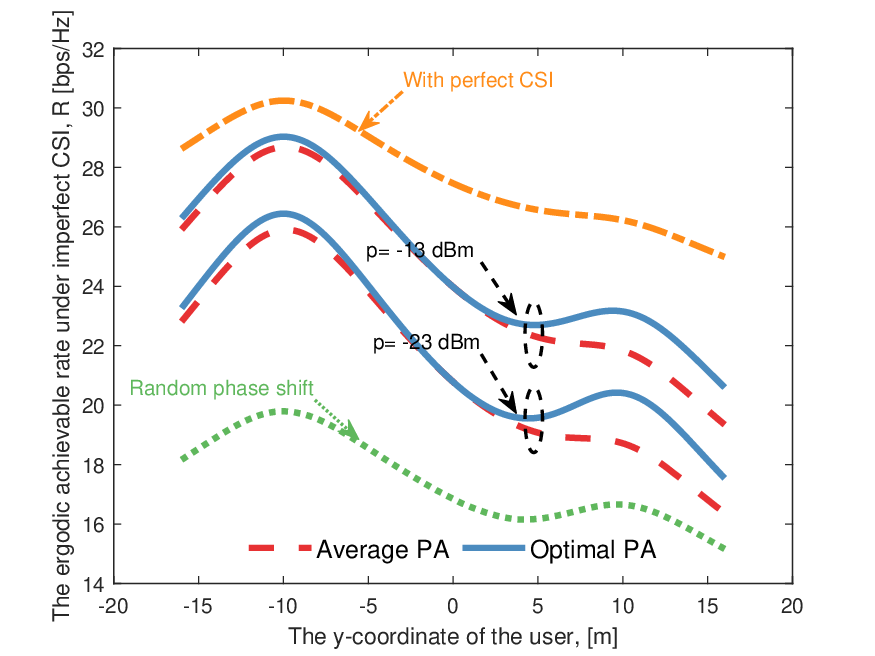}%
\label{fig8}}
\subfloat[The power of pilot signals versus the $y$-coordinate of the user.]{\includegraphics[width=8.5cm]{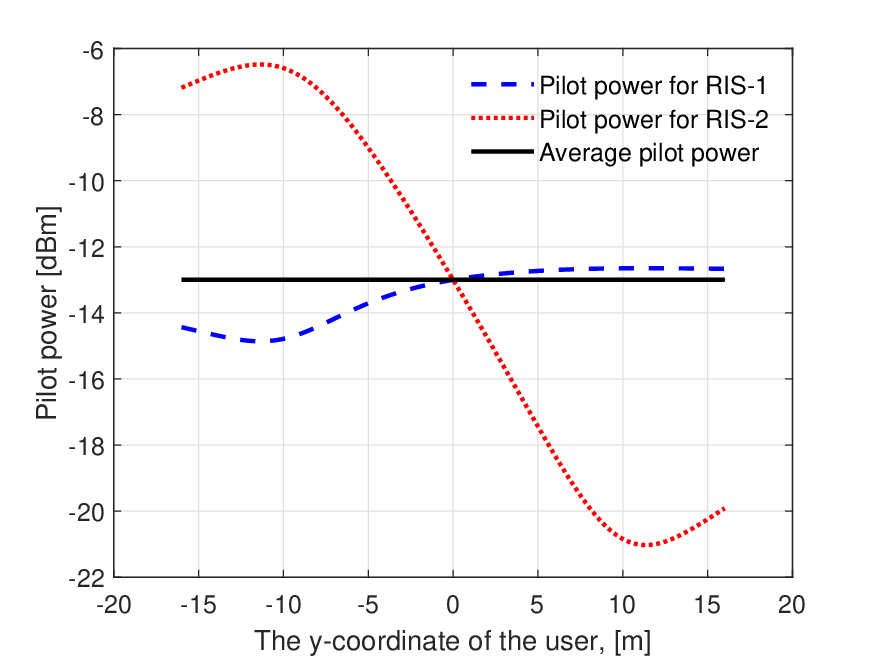}%
\label{fig9}}
\caption{Asymmetrical arrangement of two RISs.}
\end{figure}
Furthermore, let us consider a more general case where RIS-$1$ and RIS-$2$ are equipped with a different number of reflecting elements due to the specific surface area limitations. In our simulations, we set the number of reflecting elements on RIS-$1$ and RIS-$2$ to ${M_1} = 1000$ and ${M_2} = 100$, respectively. The simulation results are shown in Fig. \ref{fig8}, which are significantly different from those in Fig. \ref{fig6}. As depicted in Fig. \ref{fig8}, the ergodic achievable rate of the user near RIS-$1$ is higher than that near RIS-$2$, due to the fact that RIS-$1$ is equipped with more reflecting elements. Moreover, compared to the average pilot power allocation scheme, the proposed pilot power allocation scheme is more effective in the vicinity of RIS-$2$, which means that the reflecting elements on RIS-$1$ play a critical role in enhancing the achievable rate of the user near RIS-$2$. However, for the user in the vicinity of RIS-$1$, the smaller number of reflecting elements on RIS-$2$ hardly contributes to the rate improvement. In a nutshell, the rate improvement caused by the proposed pilot power allocation is more effective in the vicinity of a small RIS.

Finally, the specific pilot power comparison for an asymmetric number of reflecting elements is shown in Fig. \ref{fig9}. Observe from Fig. \ref{fig9} that the qualitative conclusion that allocating more power to the pilots for estimating the weak reflection channels still holds. Besides, since the number of reflecting elements on RIS-$1$ is much larger than that at RIS-$2$, makes the pilot power for estimating the reflection channels via RIS-$1$ is similar to the average pilot power allocation. Note that under all setups considered, the ratio between the highest and lowest pilot power is less than $15$ dB, which is within the dynamic range that most commercial power amplifiers can achieve.

\section{Conclusions}\label{sec7}
In this paper, we proposed the optimal pilot power allocation scheme for multi-RIS-aided communication systems. Specifically, we first derived the ergodic channel gain under imperfect CSI. The pilot power allocation problem was formulated and then solved by leveraging the method of Lagrange multipliers to maximize the ergodic channel gain. Extensive simulation results verified the performance improvements of the proposed pilot power allocation solution and provided some useful insights. Specifically, allocating more power to the pilots used for estimating the weak channel is capable of improving the ergodic channel gain, thereby enhancing the ergodic achievable rate.
\bibliographystyle{IEEEtran}
\bibliography{ref}
\end{document}